\documentclass[aps,prl,twocolumn,showpacs,superscriptaddress]{revtex4-1}

\usepackage[dvipdfm]{graphicx}
\usepackage[dvipdfm]{hyperref}
\usepackage{bm}

\begin{document}

\title{5-mg suspended mirror\\ driven by measurement-induced back-action}

\author{Nobuyuki Matsumoto}
  \email{matsumoto@granite.phys.s.u-tokyo.ac.jp}
  \affiliation{Department of Physics, University of Tokyo, Bunkyo, Tokyo 113-0033, Japan}
\author{Kentaro Komori}
  \affiliation{Department of Physics, University of Tokyo, Bunkyo, Tokyo 113-0033, Japan}
\author{Yuta Michimura}
  \affiliation{Department of Physics, University of Tokyo, Bunkyo, Tokyo 113-0033, Japan}
\author{\\Gen Hayase}
  \affiliation{Department of Chemistry, Kyoto University, Sakyo, Kyoto 606-8502, Japan}
\author{Yoichi Aso}
  \affiliation{National Astronomical Observatory of Japan
2-21-1, Osawa, Mitaka, Tokyo, Japan\\
Department of Astronomical Science, Graduate University for Advanced Studies
2-21-1, Osawa, Mitaka, Tokyo, Japan}
\author{Kimio Tsubono}
  \affiliation{Department of Physics, University of Tokyo, Bunkyo, Tokyo 113-0033, Japan}

\date{\today}

\begin{abstract}
Quantum mechanics predicts superposition of position states even for macroscopic objects.  
Recently, the use of a quasi-freely suspended mirror combined with laser was proposed to prepare such states, by M\"uller-Ebhardt et al. [Phys.Rev.Lett.{\bf100}, 013601 (2008)]. %\cite{PhysRevLett.100.013601}. 
One of the key milestones  towards this goal is the preparation of the mechanical oscillator mainly driven by quantum back-action. %, which identifies the connection between the object and quantumness of the light. 
Here, we describe development of a suspended 5-mg mirror driven by quantum back-action larger than thermal fluctuating force by a factor of 1.4$\pm0.2$ at 325 Hz, which is confirmed by using a triangular optical cavity. 
\end{abstract}

%Mechanical effects of light on material media, microstructures and particles , 42.50.Wk
%Quantum fluctuations, quantum noise, and quantum jumps, 42.50.Lc
%Cavity quantum electrodynamics; micromasers, 11.30.Cp
%Resonators, laser, 42.60.Da

\maketitle

%%%%% ===== INTRODUCTION =====
{\it Introduction.}---Recent advances in technology have enabled experimental demonstration of quantum interference using molecules\cite{arndt1999wave,gerlich2011quantum}.
However, superposition of positions of macroscopic objects has not been observed, even though quantum mechanics predicts it. 
This is at the heart of the so-called ``measurement problem".
Until now, intense works have revealed that environment such as a thermal bath plays an important role in decoherence -- the loss of quantum interference\cite{RevModPhys.75.715,hackermuller2004decoherence}.
The general relativity, on the other hand, suggests that gravity might prohibit the superposition of massive objects due to a fluctuating space-time\cite{*[{For a review, see, for example: }] [{}] RevModPhys.85.471}. 
If we prepare a massive object isolated enough from the other environment so that the superposition of position states is expected to be generated from the viewpoint of the quantum mechanics, we can test the gravity-induced decoherence.\par

For this purpose, utilization of optomechanical oscillators combined with light such as 
suspended mirrors (e.g., gravitational wave detectors such as LIGO\cite{harry2010advanced} can be used\cite{PhysRevA.81.012114}) has been proposed, because even a massive suspended mirror is expected to be entangled with the intense laser field; therefore the resulting entanglement causes the position of the oscillator to be superposed. 
However, at macroscopic scales, it becomes sensitive to environmental disturbances because it is difficult to isolate the oscillator from the environment.
Indeed, quantum back-action derived from Heisenberg's uncertainty principle\cite{heisenberg1927} has been observed using nano/micro mechanical oscillators under Planck mass ($\sim$ 22 ${\rm \mu g}$)\cite{murch2008observation,PhysRevLett.108.033602,PhysRevA.86.033840,Purdy15022013}, while it is generally masked by thermal noise in the case of using massive oscillators like a suspended mirror. 
This is partially due to a technical limitation -- the radiation pressure of light will expose a free mass to instability such as anti--torsional spring effect\cite{Sidles2006167}.
Furthermore, there is a fundamental compromise between tolerance for the instability and sensitivity\cite{Matsumoto:14}; sufficient tolerance with firm suspension makes the mass differ from free mass, and this results in increase of a thermal fluctuating force. 
In optomechnical systems, quantum back-action is induced by quantum fluctuations of the probe light, and the oscillator driven by its fluctuation is necessary for avoiding the decoherence processes, e.g., due to suspension thermal noise or seismic noise.
Here, we continuously measured the position of a suspended 5-mg mirror and the force imposed on it using a triangular optical cavity, and satisfying this condition.\par 

\begin{figure}
\begin{center}
\includegraphics[scale=0.45]{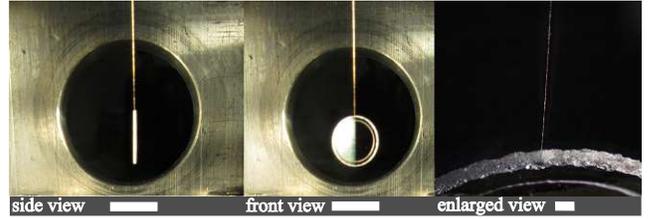}
\end{center}
\caption{Mechanical oscillator.
The mirror was manufactured by SIGMA KOKI.
It has radius of 2\ mm, thickness of 0.2\ mm, and mass of 5\ mg.
The tungsten wire of 3\ ${\rm \mu m}$ diameter and 50\ mm length is attached on the mirror with epoxy resin.
Both in the side view and the front view, the tungsten wire appears much thicker than the actual size because of the overexposure of the camera.
The enlarged view photographed by a stereoscopic microscope (Olympus, SZ61) shows the interface between the wire and the mirror.
Scale bars, 4\ mm both in side and front view, and 0.2\ mm in enlarged view.}
\label{fig1}
\end{figure}

%%%%% ===== TRIANGULAR OPTICAL CAVITY =====
{\it Triangular optical cavity.}---There are two main technical features: an extremely thin suspension wire and triangular geometry of the cavity. 
Firstly, the thin wire assures that the amount of energy stored in the pendulum is dominated by the gravitational potential, and thus the mechanical loss of the pendulum, $\gamma_{\rm pend}$, is diluted by a factor of $k_{\rm grav}/k_{\rm el} (\propto 1/r^2)$\cite{PhysRevD.42.2437}. 
Here, $k_{\rm grav}$ and $k_{\rm el}$ are respectively the gravitational and elastic spring constants of the pendulum, and $r$ is the radius of the wire. 
The use of a thin wire decreases the mechanical loss; however, it is weak for an optical torsional anti-spring effect\cite{Sidles2006167}, due to the low mechanical restoring force of the wire.  
Use of a triangular cavity overcomes this limitation due to an optical torsional positive-spring effect\cite{Matsumoto:14}. 
In our setup, the optical torsional spring constant is $k_{\rm t,opt}=1.2\times10^{-9}$\ Nm/rad whereas it is $k_{\rm t,opt}^{(\rm linear)}=-k_{\rm t,opt}$ for a linear cavity with otherwise the same scale and power. 
We succeed in storing 51 times higher optical power (4.1 W) than the instability limit for the linear cavity, while the dilution factor becomes about 600.\par

\begin{figure}
\begin{minipage}{0.95\hsize}
  \begin{center}
   \includegraphics[width=82mm]{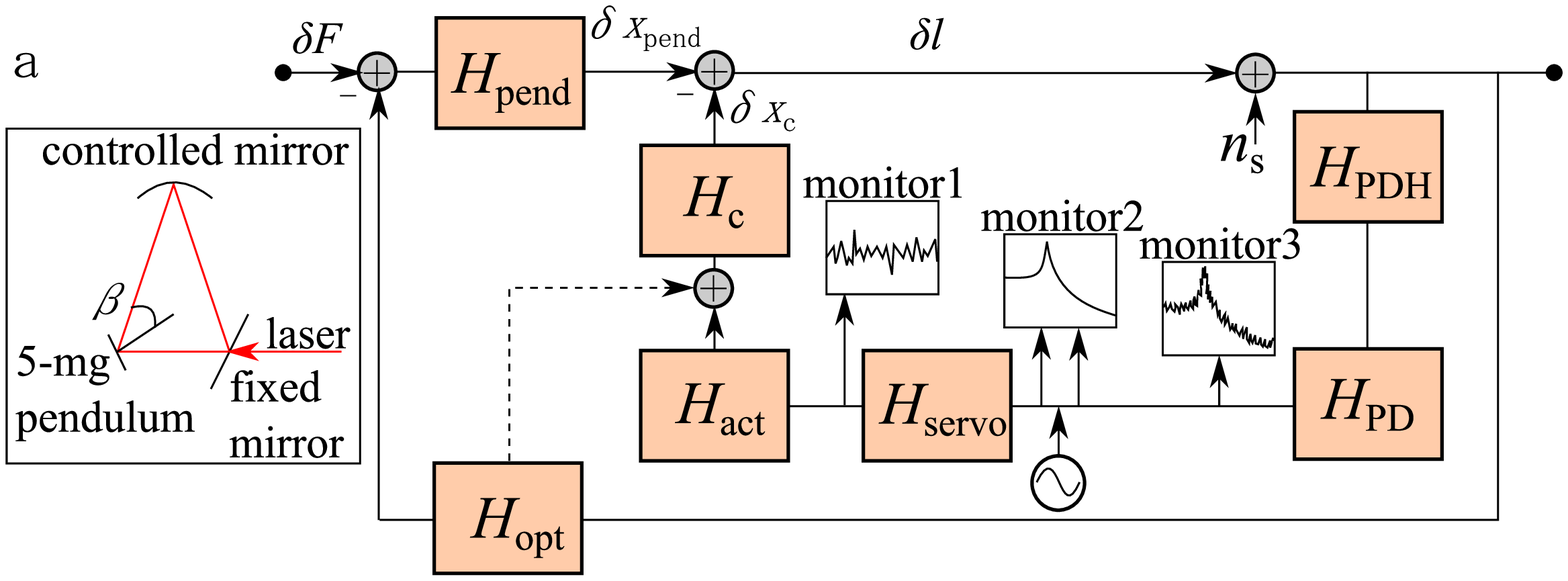}
  \end{center}
 \end{minipage}
 \begin{minipage}{0.49\hsize}
  \begin{center}
   \includegraphics[width=44mm]{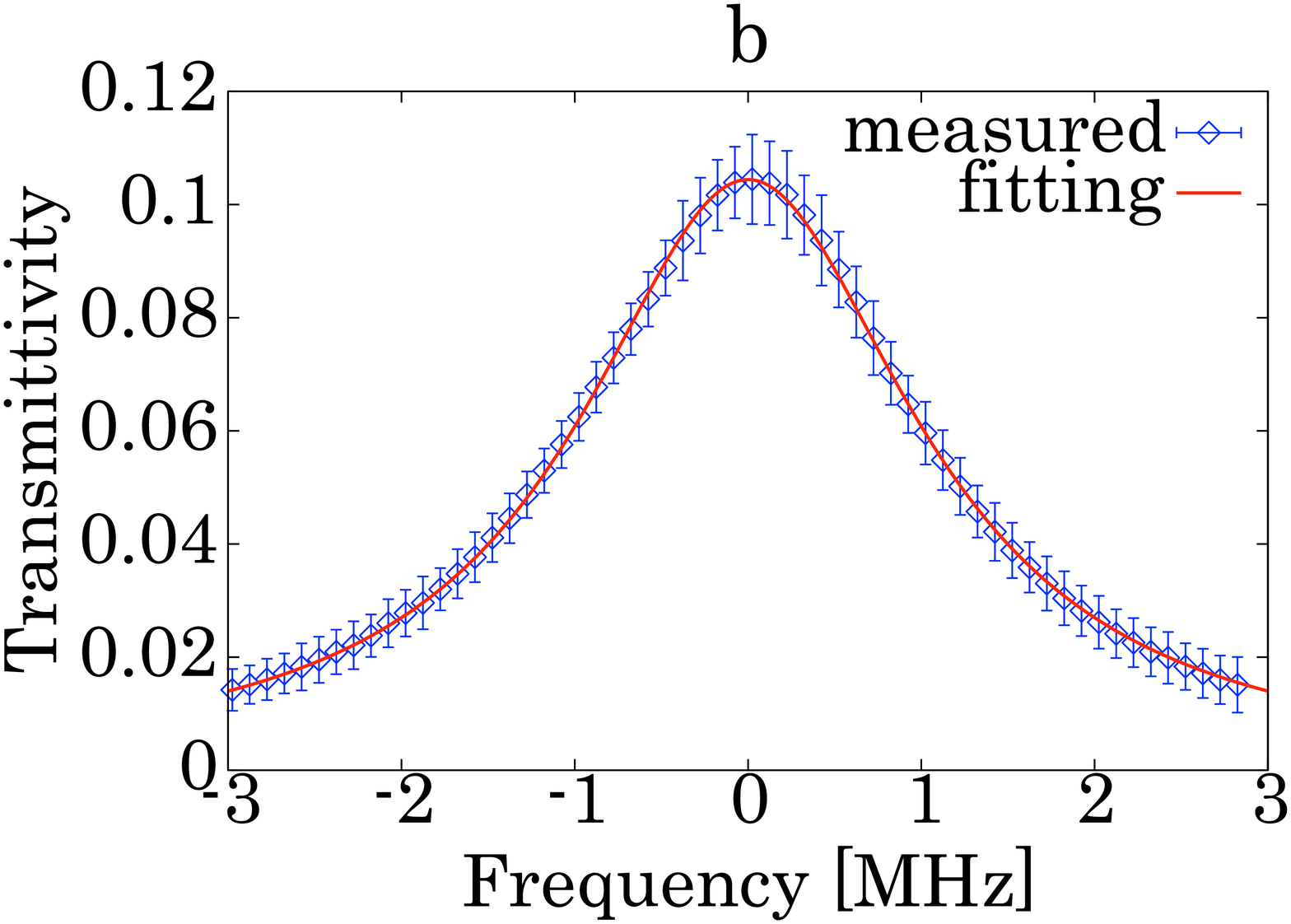}
  \end{center}
 \end{minipage}
 \begin{minipage}{0.49\hsize}
  \begin{center}
   \includegraphics[width=44mm]{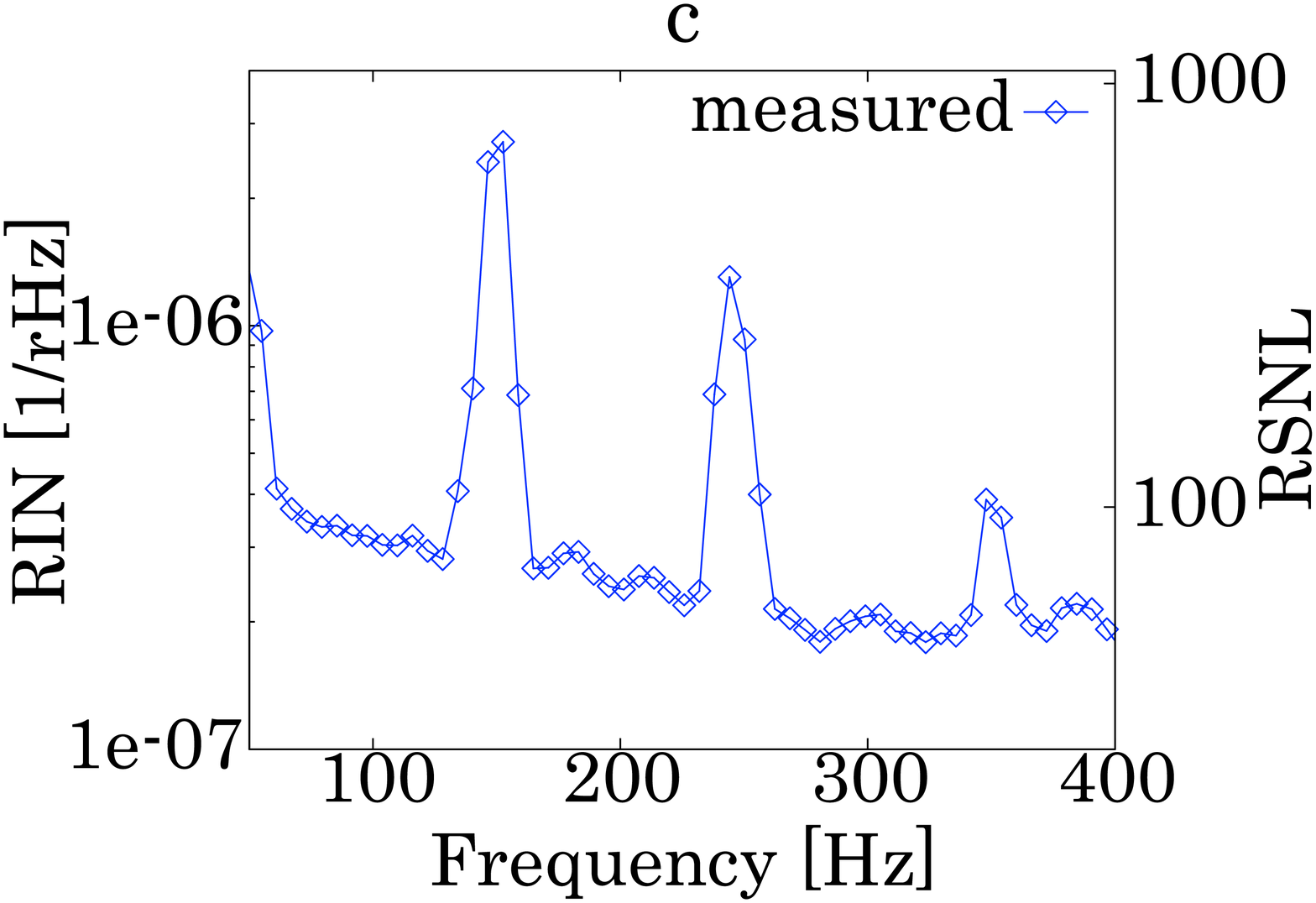}
  \end{center}
 \end{minipage}
\begin{minipage}{0.49\hsize}
  \begin{center}
   \includegraphics[width=44mm]{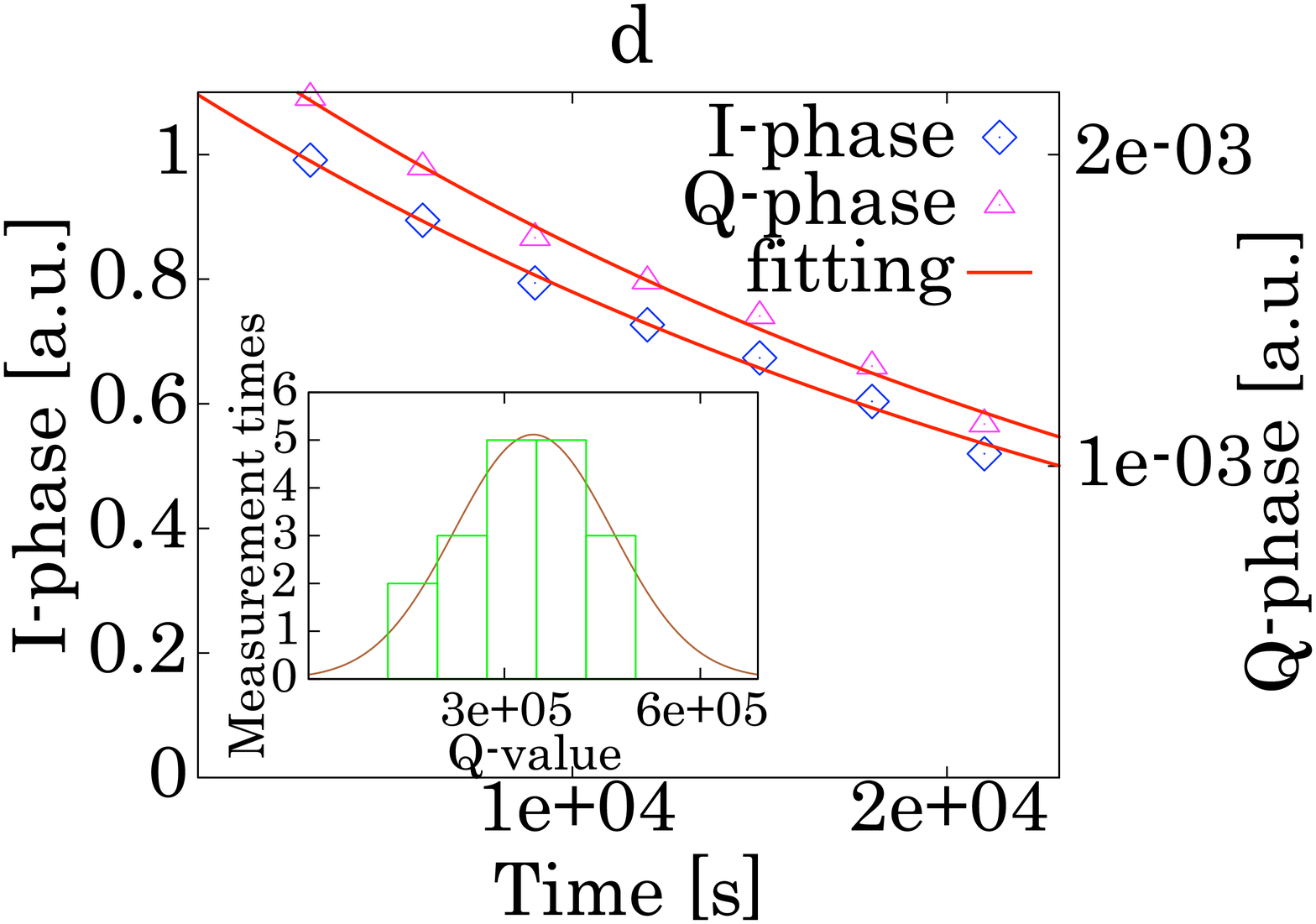}
  \end{center}
 \end{minipage}
\begin{minipage}{0.49\hsize}
  \begin{center}
   \includegraphics[width=44mm]{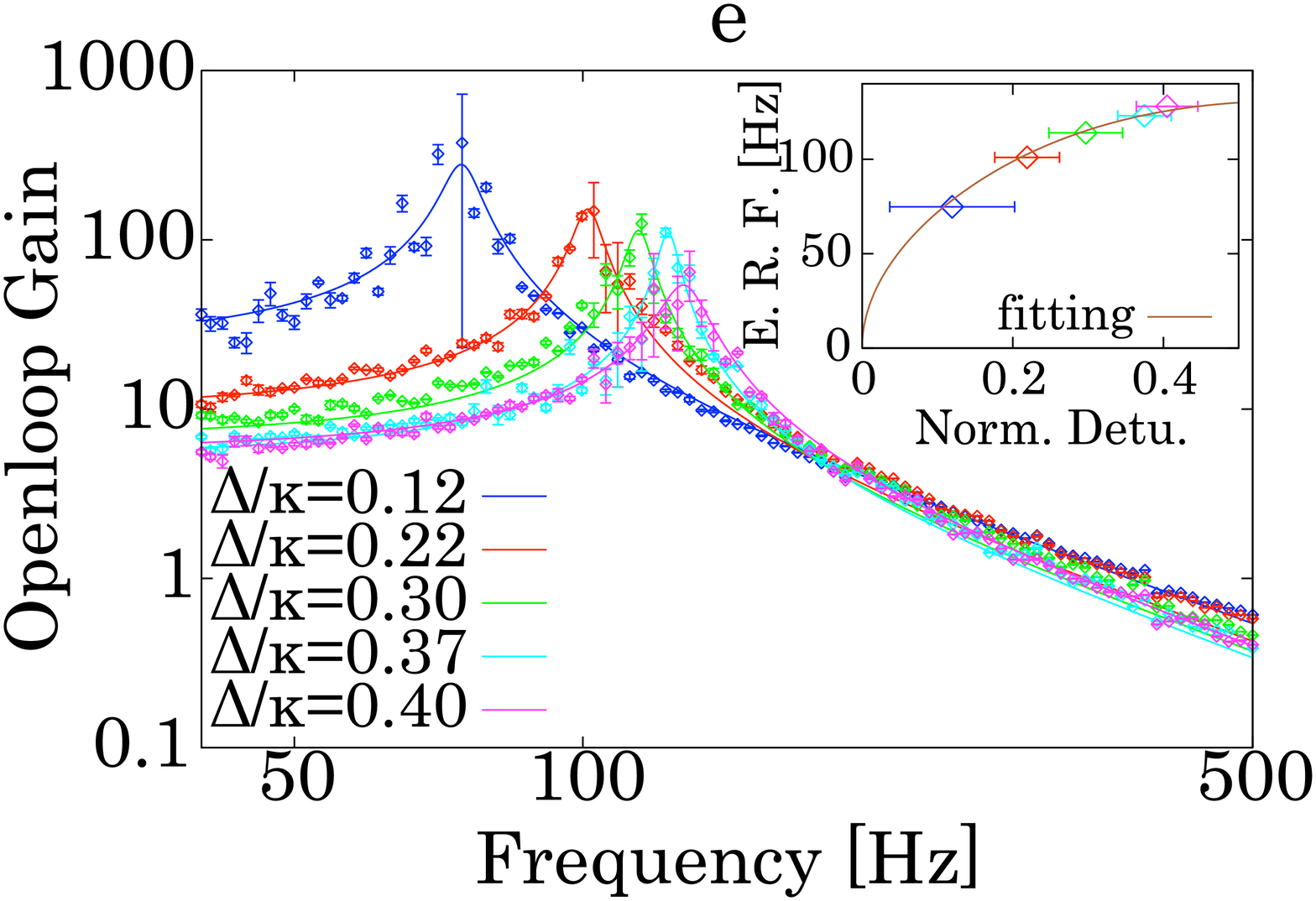}
  \end{center}
 \end{minipage}
 \caption{{\bf a:} Block diagram of the experiment. 
Force fluctuation imposed on the 5-mg pendulum is respectively monitored as force and displacement power spectrum at the monitor1 and the monitor3. 
The dashed line shows the optical spring effect to the controlled mirror, but it is negligible because the controlled mirror is too massive to be moved by the optical force. 
The laser is incident from the fixed mirror (incident angle given by $\beta$ is estimated to be $0.75\pm0.04$ rad) as shown in the inset figure. 
{\bf b:} The optical cavity linewidth. 
{\bf c:} The measured spectra of the intensity fluctuation of the input laser. 
Left vertical axis indicates the relative intensity noise level (RIN), and the right side indicates the relative shot noise level (RSNL).
Spectral peaks are identified as power line harmonic. 
{\bf d:} The ringdown of the pendulum. 
Amplitude (I-phase) and phase quadrature (Q-phase) of the oscillation during a free decay are obtained by the optical shadow sensor. 
The inset shows the distribution of the measured mechanical quality factor, and is fitted by the Gaussian distribution. 
{\bf e:} The optical spring effect. 
The effective mechanical frequency is measured by monitoring the open-loop function given by $G_2/(1-G_1)$ at the monitor2. 
The inset shows the dependence of the effective resonant frequency (E. R. F.) on the cavity detuning normalized by the cavity linewidth.}
\label{fig2}
\end{figure}

%%%%% ===== EXPERIMENT =====
{\it Experiment.}---Our optomechanical system is a triangular cavity, which consists of the suspended 5-mg mirror (Fig. \ref{fig1}), a fixed mirror, and a suspended $1\times10^2$-g mirror with actuators attached for cavity length control, whose mass is determined so that cavity length fluctuation due to the thermal motion of this controlled mirror is negligibly small. 
The 5-mg mirror is suspended by a tungsten wire of 50\ mm length with 3\ ${\rm \mu}$m diameter attached to the mirror with epoxy resin. 
These are placed on a vibration isolation stage installed in a vacuum chamber ($1\times10^{-3}$ Pa), whose pressure is determined such that the residual gas damping is negligible. 
The cavity is controlled on its resonance (or slightly detuned for generating the optical spring\cite{PhysRevA.69.051801}) by the Pound-Drever-Hall (PDH) method\cite{Drever1983} using a Nd:YAG laser source with a wavelength of 1064 nm, as shown in Fig. \ref{fig2}a.
We measure two types of the back-action signal: the displacement power spectrum from the PDH signal with no feedback gain, i.e. the controlled $1\times10^2$-g mirror is {\it not} electrically trapped such that the displacement signal can be directly measured; and the force power spectrum from the PDH signal with some gain, i.e. the controlled mirror is electrically trapped with respect to the cavity length such that the controlled mirror works as a transducer of the force fluctuation acting on the 5-mg mirror.\par

The detailed expression is given by using follow parameters shown in Fig. \ref{fig2}a, 
$\delta F$ [N], force fluctuation imposed on the 5-mg pendulum; 
$\delta l$ [m], displacement fluctuation of the cavity length; 
$\delta x_{\rm pend}$ [m], displacement of the pendulum; 
$\delta x_{\rm c}$ [m], displacement of the controlled mirror; 
$H_{\rm PDH}$ [W/m], power-to-displacement conversion factor; 
$H_{\rm PD}$ [V/W], voltage-to-power conversion factor; 
$H_{\rm servo}$ [V/V], servo filter; 
$H_{\rm act}$ [N/V] (it is measured to be $(2.1\pm0.1)\times10^{-5}$ N/V by using a simple Michelson interferometer), the efficiency of the actuator; 
$H_{\rm pend}$ [m/N], mechanical susceptibility of the pendulum; 
$H_{\rm c}$ [m/N], mechanical susceptibility of the controlled mirror; 
$H_{\rm opt}$ [N/m], the optical spring effect; and
$n_{\rm S}$ [m], the sensing noise.
The monitor1 gives the force fluctuation as $G_2H_{\rm pend}/(H_{\rm c}H_{\rm act})\times(\delta F+n_{s}/H_{\rm c})/(1-G_1-G_2)$ [V] where $G_1=H_{\rm pend}H_{\rm opt}$ and $G_2=H_{\rm PDH}H_{\rm PD}H_{\rm servo}H_{\rm act}H_{\rm c}$, while the monitor3 gives the displacement fluctuation as $(\delta x_{\rm pend}+n_{\rm s})H_{\rm PDH}H_{\rm PD}$ [V] where $G_2$ is supposed to be negligibly small. 
Force spectrum differs from displacement spectrum in that it is not be affected by the changeable susceptibility of the pendulum due to the dynamical back-action of the optical field, as it is independent on the mechanical dynamics of the pendulum. 
Displacement spectrum, on the other hand, gives us the information of the mechanical susceptibility, and thus is tolerant of sensing noise--dummy signal that limits the sensitivity for measurement, which is independent from mechanical motion-- at the resonance. \par

\begin{figure*}
\begin{minipage}{0.48\hsize}
\includegraphics[width=65mm]{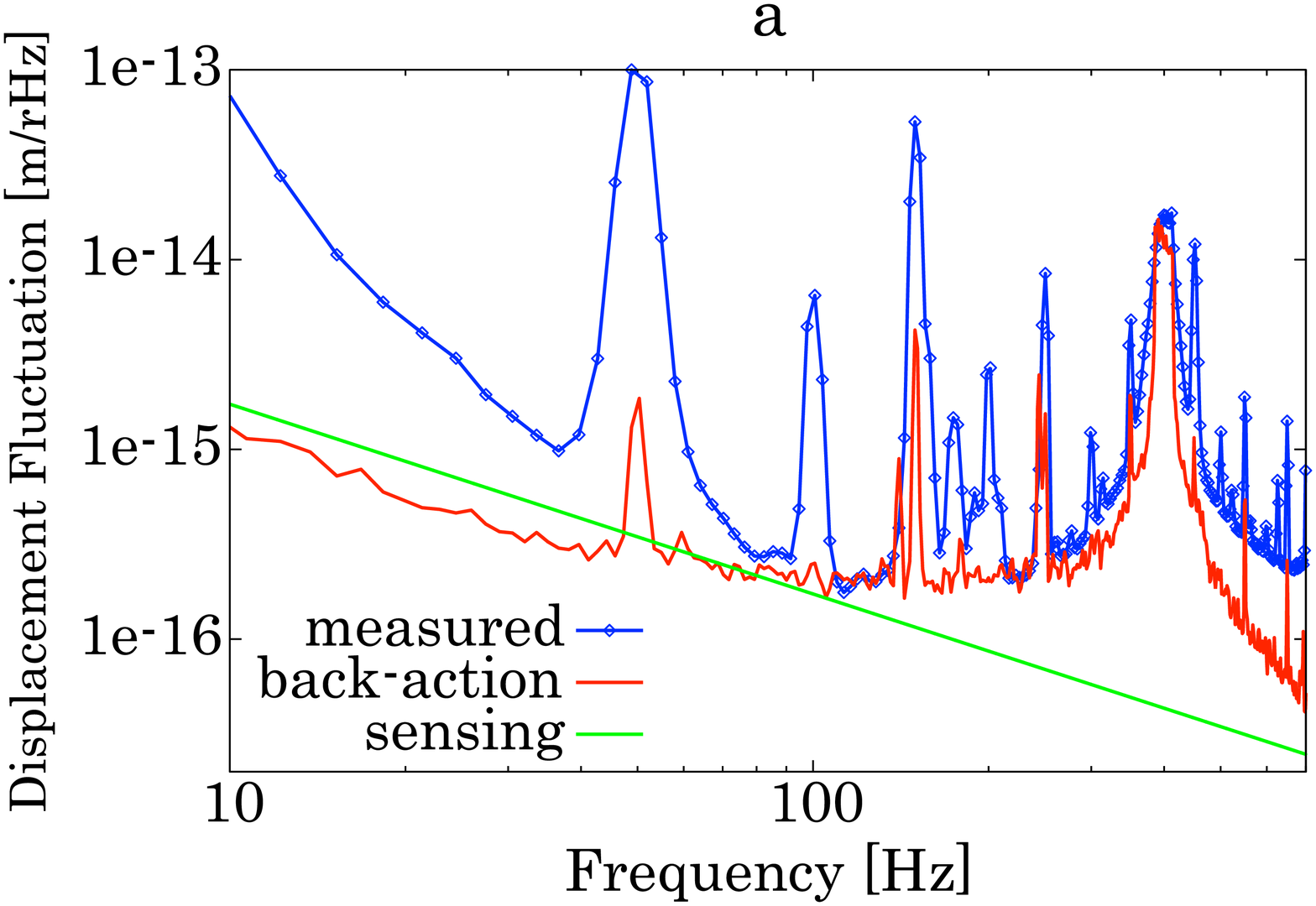}
\centering
\end{minipage}
\begin{minipage}{0.48\hsize}
  \begin{center}
   \includegraphics[width=65mm]{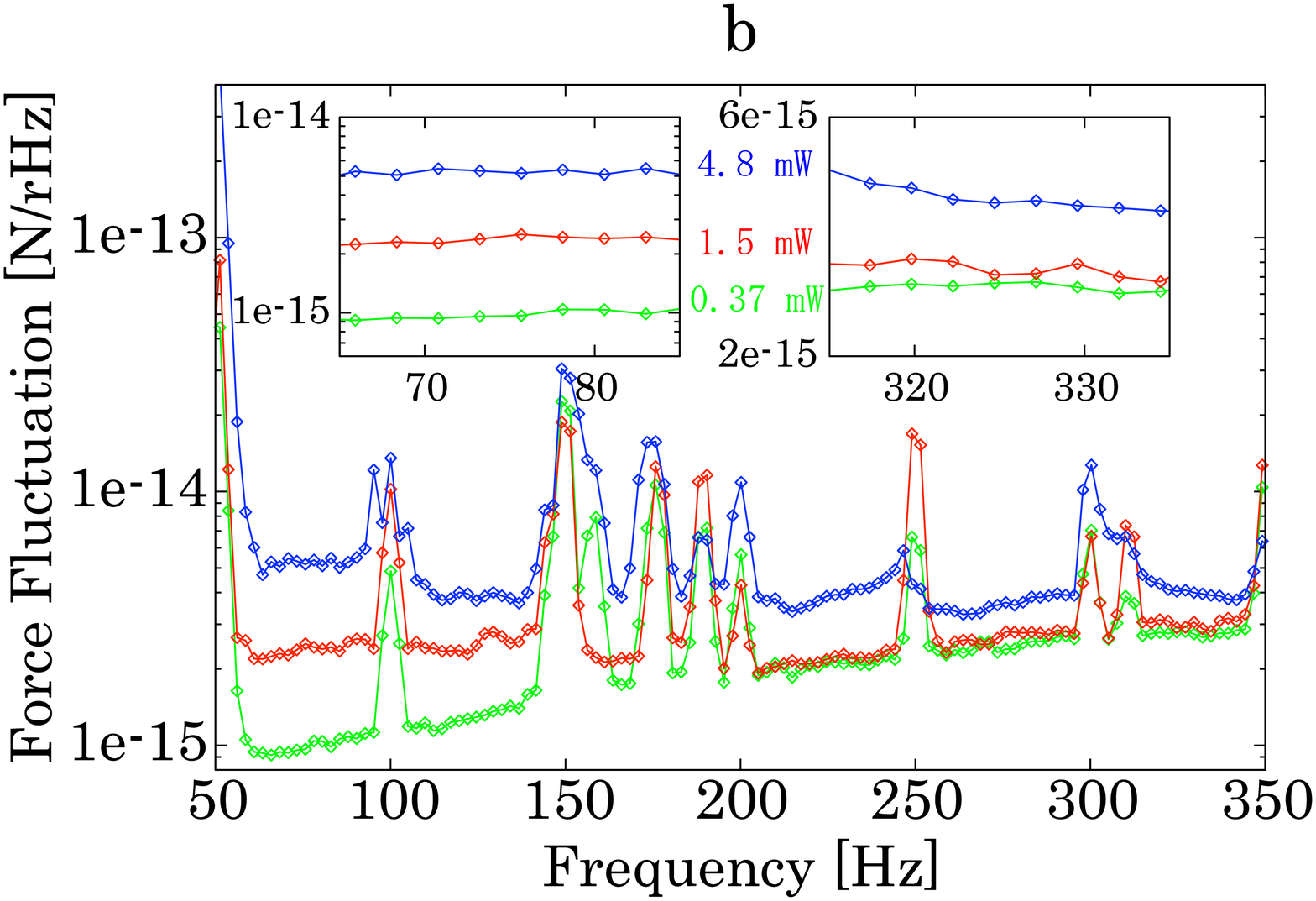}
  \end{center}
 \end{minipage}
\caption{
{\bf a:} Observed spectra of displacement fluctuation at optical power, $P_{\rm in}=7.6\ {\rm mW}$; cavity detuning, $\Delta=1.1\times\kappa$; and open loop gain, $G_2=0$.
Measured displacement spectral density (blue), estimated back-action contribution (red), and estimated sensing noise with $f^{-1}$ ($f^1$ in the force fluctuation) spectral slope (green) are shown. 
Spectral peaks are identified as follows: at around 200 Hz, suspension wire violin mode; at around 400 Hz, pendulum motion trapped by the optical spring. 
{\bf b:} Observed spectra of force fluctuation (blue: input laser power of 4.8 mW; red: 1.5 mW; green: 0.37 mW). }
\label{fig3}
\end{figure*}

To study the force fluctuation imposed on the pendulum accurately, we make a series of measurements to characterize our optical, mechanical, and opto-mechanical systems. 
As shown in Fig. \ref{fig2}b-e, the half linewidth of the cavity, $\kappa$, is measured to be $2\pi\times(1.181\pm0.003)$\ MHz by sweeping the laser frequency across the optical resonance; intensity fluctuations of the input beam normalized by its power, $A$, is measured to be $3.5\times10^{-7} /\sqrt{\rm Hz}$ at 75 Hz (the relative shot noise level in amplitude, $B=\sqrt{\rho P_{\rm in}/(2e)}A$, is estimated to be 94, where $e$ is elementary charge, $\rho (=0.73\pm0.07\ [{\rm A/W}])$ is the quantum efficiency, and $P_{\rm in}$ is input laser power) by the direct photo-detection; the mechanical quality factor of the 5-mg pendulum, $Q_{\rm pend}$, is measured to be $(3.2\pm1.0)\times10^5$ by the ring-down measurement; and the single-photon optomechanical coupling rate, $g$,  is measured to be $2\pi\omega_{\rm c}\times(2.8\pm0.2)$/m, where $\omega_{\rm c}$ is the cavity resonant frequency, by observing the optical restoring force. 
By using these parameters, (double-sided) power spectrum of force fluctuation induced by the quantum back-action, the back-action due to the classical intensity fluctuation, and the thermal bath are respectively given by $S_{\rm FF,q}^{(2)}=2N_{\rm circ}\hbar^2g^2/\kappa$, $S_{\rm FF,c}^{(2)}=4\kappa_{\rm in}/\kappa\times B^2\times S_{\rm FF,q}^{(2)}$, and $S_{\rm FF,th}^{(2)}=4k_{\rm B}T\gamma_{\rm pend}m$. 
Here, $\hbar$ is the reduced Planck constant, $N_{\rm circ}$ is mean photon number inside the cavity, $T$ is temperature of the pendulum, and $m$ is the effective mass of the pendulum. 
We note that no one knows whether the dissipation is depend (structure damping) or independent (viscous damping) on the frequency {\it a priori}, and thus the ratio of the quantum back-action to the thermal fluctuating force becomes $S_{\rm FF,q}^{(2)}/S_{\rm FF,th}^{(2)}=(N_{\rm circ}g^2/n_{\rm th}\kappa)\times(2Q/\omega_{\rm m})$ or $(N_{\rm circ}g^2/n_{\rm th}\kappa)\times(2Q\omega/\omega_{\rm m}^2)$, where $n_{\rm th}$ is phonon number, $\omega_{\rm m}/2Q$ is the mechanical dissipation for the viscous case, and $\omega_{\rm m}^2/2Q\omega$ is for the structure case.  
Because our pendulum can be trapped by the optical spring, the ratio at the pendulum resonant frequency can be further increased with increased optical restoring force by a factor of $\omega_{\rm eff}/\omega_{\rm m}$, if the dissipation of the pendulum is limited by the internal friction, i.e. the structure damping.  
Also, the back-action induced by the phase fluctuation (labeled by $\delta\phi$) of the laser\cite{PhysRevA.80.063819}, which is written by $S_{\rm FF,phase}^{(2)}=S_{\rm FF,c}^{(2)}\times2\Delta\omega_{\rm eff}\delta\phi/(\kappa^2+\Delta^2)$, is negligible because the cavity has a large linewidth than the effective resonant frequency of the pendulum (labeled by $\omega_{\rm eff}$), whose condition is generally called ``bad" cavity condition\cite{RevModPhys.86.1391}. 
In our measurements, roughly only 0.3\% of the force fluctuation is due to the phase noise. \par

\begin{figure}
 \begin{minipage}{1\hsize}
  \begin{center}
   \includegraphics[width=88mm]{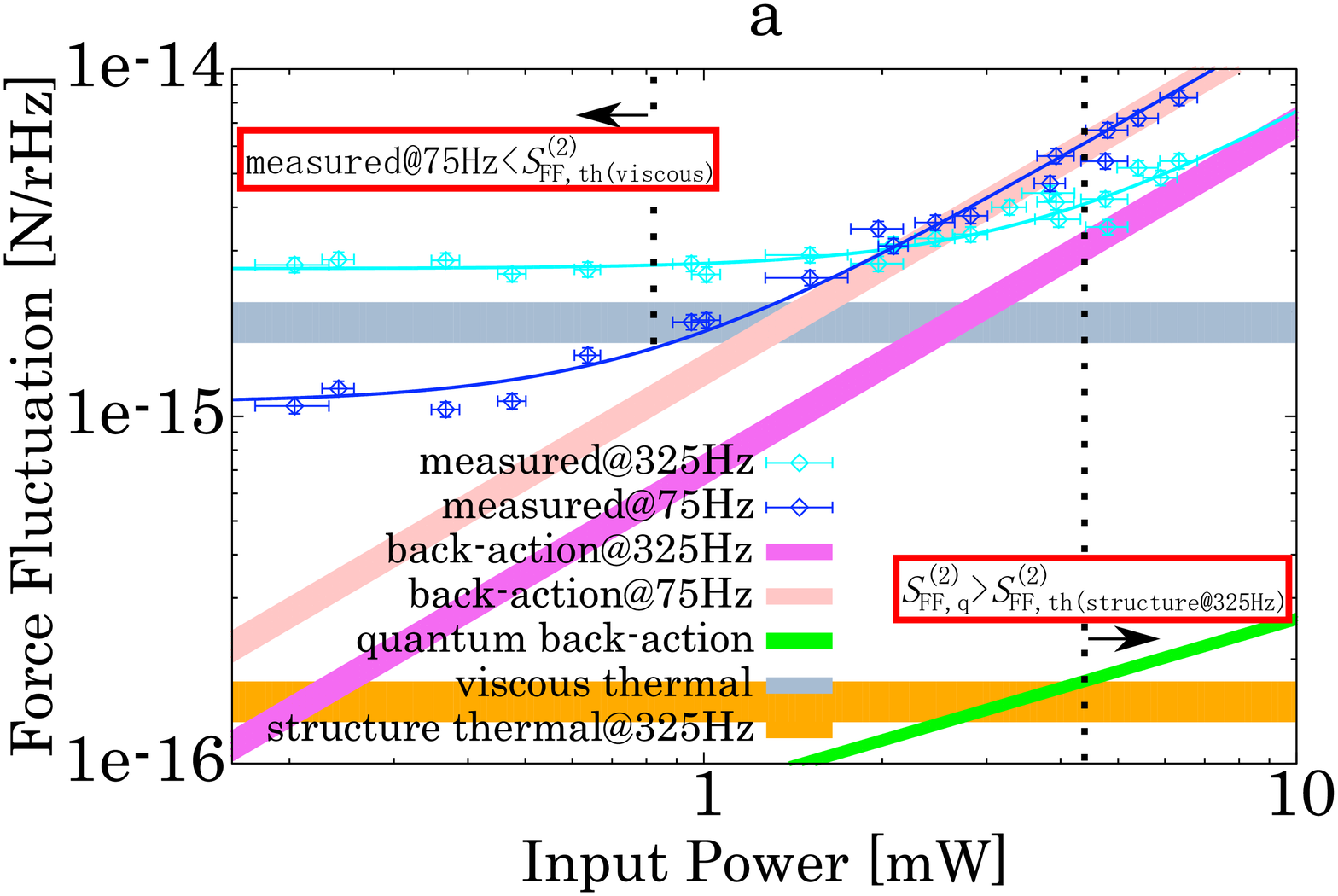}
  \end{center}
 \end{minipage}
\begin{minipage}{0.49\hsize}
  \begin{center}
   \includegraphics[width=44mm]{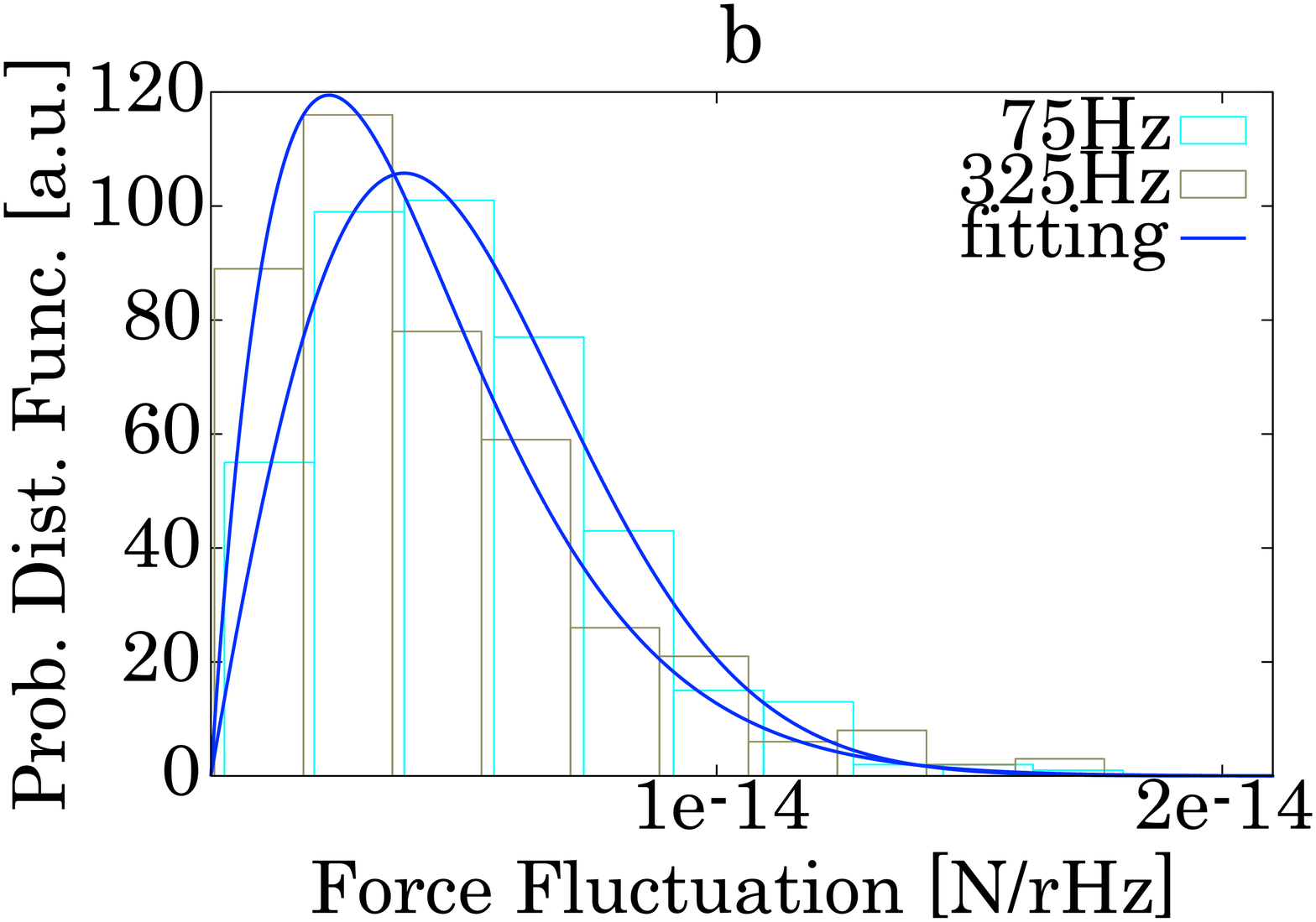}
  \end{center}
 \end{minipage}
 \begin{minipage}{0.49\hsize}
  \begin{center}
   \includegraphics[width=44mm]{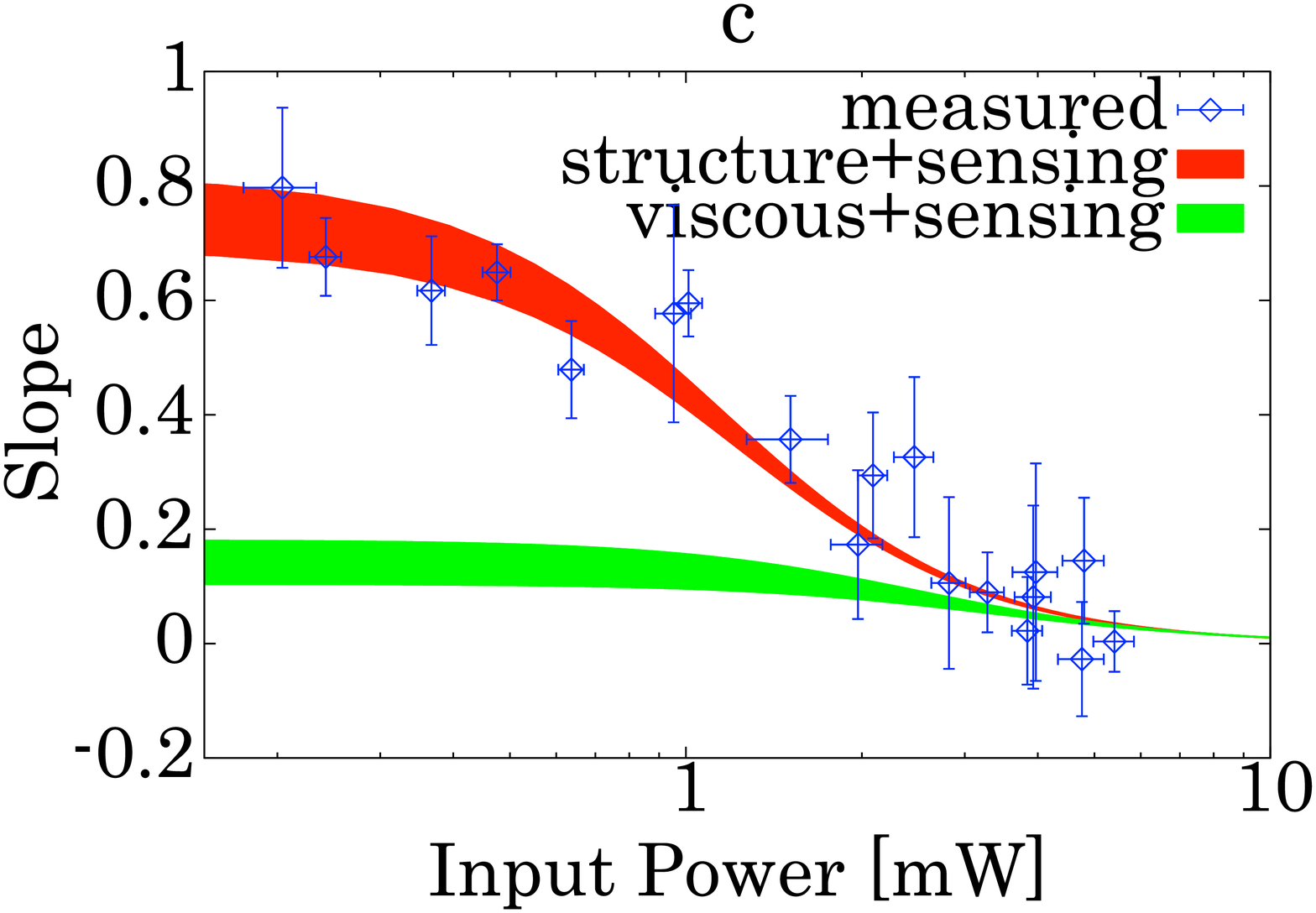}
  \end{center}
 \end{minipage}
 \caption{{\bf a:} Dependence of the measured force amplitude spectral density on the input laser power.
Measured force fluctuation at 325 Hz (blue), measured force fluctuation at 75 Hz (cyan), theoretical thermal force spectrum with the viscous damping model (gray), theoretical thermal force spectrum with the structure damping model at 325 Hz (orange), and estimated back-action with the measured $B=94 (A=3.5\times10^{-7} /\sqrt{\rm Hz})$ at 75 Hz and $B=48 (A=1.8\times10^{-7} /\sqrt{\rm Hz})$ at 325 Hz (light red and magenta) are shown. 
Each area includes the 68\% confidence level. 
The error is due to the systematic error such as the uncertainty of the quantum efficiency and calibration factor, and the statistic error in measurement. 
{\bf c:} Distribution of force amplitude spectral density with input power of 4.8 mW for two representative frequencies within the measurement band. 
Each curve is a histogram of the spectrum at the specified frequency. 
Each of them is taken from the Fourier transform of 0.4 s of data; the equivalent noise bandwidth for each curve is 2.4 Hz. 
{\bf d:} Dependence of the spectrum slopes at 75 Hz on the input laser power.
Measured slopes of the force amplitude spectral density (blue), theoretical estimation based on the structure damping model (red), and the theoretical estimation based on the viscous damping model (green) are shown.   }
\label{fig4}
\end{figure}

%%%%% ===== DATA ANALYSIS =====
{\it Data analysis.}---The measured (single-sided) amplitude spectral density (the square root of the power spectrum) of the optically trapped pendulum motion, with the input power of 7.6 mW and the cavity detuning of $1.1\times\kappa$, is shown in Fig. \ref{fig3}a as blue dots. 
The calibrated noise level agrees with the estimated motion by the back-action (red line) at around 400 Hz, where the pendulum motion is included. 
The dependence of the measured force fluctuation shown in Fig. \ref{fig3}b at 75 and 325 Hz on the input laser power are respectively shown in Fig. \ref{fig4}a as blue and cyan dots. 
The results are also well fitted to the estimated dependence on the power over 2 mW (pink) and 5 mW (magenta) respectively, while the noise level below 0.8 mW at 75 Hz is clearly lower than the estimated thermal noise with the viscous model (gray). 
To guarantee the stationarity of our measurement, the chi square test was used to test whether a set of data fits a Rayleigh distribution. 
As shown in Fig. \ref{fig4}b, e.g., each curve is well close to Rayleigh distributions written by green lines, since they exhibited stationary. 
To distinguish the measured noise below 0.8 mW from the thermal noise with the viscous model, Fig. \ref{fig4}c shows the dependence of the spectral slope on the input laser power. 
The result is well fitted to the structure model with $f^{-1}$ spectral slope plus unknown noise with $f^1$ slope written as red area. 
Because the rms force noise must not diverge, stationary force noise has spectral slope smaller than 0; therefore roughly 93\% of the measured spectrum with no input-laser at 75 Hz is not force but sensing noise, with about $f^{1}$ spectral slope (e.g., frequency noise has such a dependence\cite{laserfreq}).
Thus the structure damping model is valid in our measurements. 
We can then estimate that the ratio of the quantum back-action to the thermal fluctuating force is larger than 1 over 325 Hz with input laser power of about 5 mW as shown in Fig. \ref{fig4}a. \par

%%%%% ===== CONCLUSION =====
{\it Conclusion.}---The geometrical advantages of the triangular cavity enables for the mirror to be isolated from the thermal bath under high intracavity power, allowing us to increase the quantum back-action under the low suspension thermal noise. 
Thus, we developed the 5-mg massive pendulum mainly driven by back-action, whose quantum component was also larger than thermal fluctuating force at room temperature. 
This is the first step toward the experimental validation of macroscopic quantum mechanics by the laser interferometer, which consists of the massive suspended mirrors. 
\acknowledgements
We thank Yutaro Enomoto, Yutaka Shikano, and Kenshi Okada for useful discussions, Shigemi Otsuka for his help.
This work was supported by the Grant-in-Aid for JSPS Fellows No. 25$\cdot$10490 and NINS Program for Cross-Disciplinary Study.  
% partially supported by Grant-in-Aid for Scientific Research (A) 22244049.


\end{document}